\documentclass{article}
\usepackage{spconf,amsmath,graphicx}
\usepackage[T1]{fontenc}

\usepackage{amsmath}
\usepackage{amssymb}
\usepackage{multirow}
\usepackage{array}
\usepackage[table]{xcolor}
\usepackage{array}

\title{ Frame-based overlapping speech detection using Convolutional Neural Networks}
\name{Midia Yousefi, John H.L. Hansen}
\address{ Center for Robust Speech Systems (CRSS), Erik Jonsson School of Engineering,\\
University of Texas at Dallas, Richardson, Texas, U.S.A.}

\begin{document}
%
\maketitle
\begin{abstract}
 Naturalistic speech recordings usually contain speech signals from multiple speakers. This phenomenon can degrade the performance of speech technologies due to the complexity of tracing and recognizing individual speakers. In this study, we investigate the detection of overlapping speech on segments as short as 25 ms using Convolutional Neural Networks. We evaluate the detection performance using different spectral features, and show that pyknogram features outperforms other commonly used speech features. The proposed system can predict overlapping speech with an accuracy of 84\% and Fscore of 88\% on a dataset of mixed speech generated based on the GRID dataset. 
\end{abstract}
\vspace{-0.2cm}
\section{Introduction}
\vspace{-0.2cm}

Spontaneous conversations such as meetings, debates, and telephone conversations tend to contain overlapping speech, i.e., time segments where more than one speaker is active \cite{adda2008annotation}. Human brain is capable of focusing on a single talker in a multi-speaker environment, recognizing both the identity of the talker and also the content of the speech. However, the performance of speech analysis technologies such as speaker diarization, identification and Automatic Speech Recognition (ASR) is adversely affected in presence of co-channel speech \cite{boakye2008overlapped,ghorbani2018advancing,jafarlou2019analyzing}. In speaker diarization, the existence of overlapping speech in the training dataset leads to generating impure speaker models, which increases diarization error \cite{knox2012did}. Also, in spite of all successful attempts in recognizing speech signals automatically, transcribing all streams of the co-channel recordings is still one of the hardest challenges in ASR systems \cite{rennie2010single,ghorbani2019domain}.

Researchers have addressed co-channel speech challenge using two major approaches; \emph{(i)} detecting the overlapping speech segments to be either removed from the dataset or to be analyzed separately for extracting useful information about the speaker identities or the speech content \cite{boakye2008overlapped}, \emph{(ii)} separating the individual speech signals from the mixture  before feeding them to the speech analysis systems. Since each of the aforementioned approaches have their own advantages and disadvantages, choosing the right one depends on the application. However, as discussed in \cite{smolenski2011usable}, for many applications, the former approach, i.e., overlapping speech detection suffices for improving the performance of the diarization/identification systems in co-channel conditions.
\vspace{-0.4cm}
\section{Related works}
\vspace{-0.3cm}
The overlapping speech detection systems can be mainly categorized into two classes: \emph{(1)} unsupervised and \emph{(2)} supervised. The former usually uses signal processing methods to design suitable features for detecting overlapping segments. In \cite{yantorno2000study}, Spectral Auto-correlation Peak Valley Ratio (SAPVR) is used to tag overlapping segments. Due to the spectrum harmonicity of the single speaker segments, the auto-correlation function tends to be periodic, however it has smaller values in the overlapping segments which can be used to manifest the presence of an interfering talker. 

Some other techniques look into the statistics of the speech signal to detect overlapping speech. Kurtosis is used in \cite{wrigley2004speech} to measure the Gaussianity of the speech segments. If there is only one active speaker in the segment, the distribution of the speech is more similar to a Gamma or Laplace distribution, however for segments with more than one active talker, the distribution tends to be more Gaussian. Therefore, kurtosis can effectively detect overlapping speech.

\begin{figure}
\begin{center}
\includegraphics[width=\linewidth]{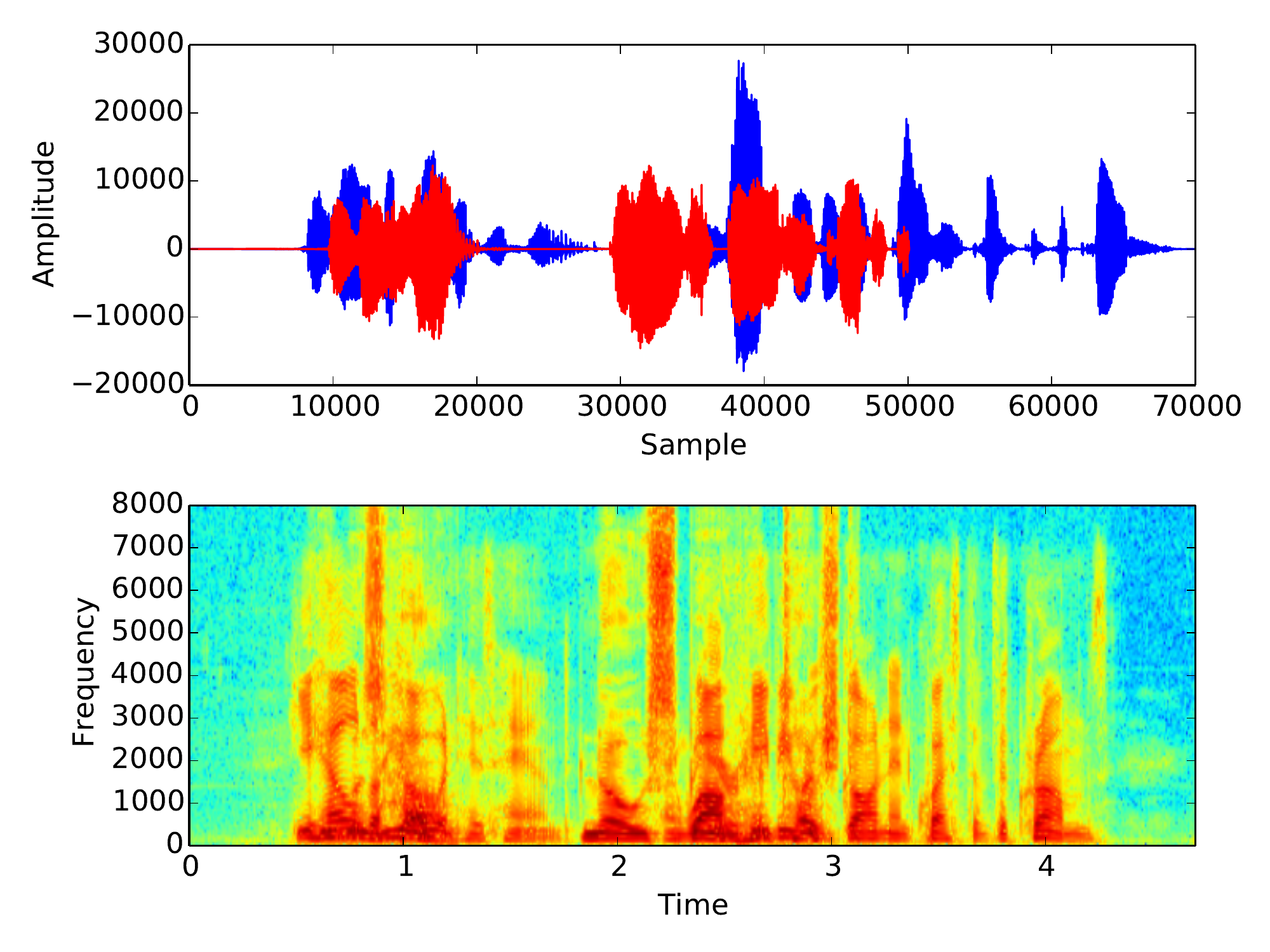}
\vspace{-0.9cm}
 \caption{Co-channel speech recording. Top: time domain waveform, bottom: frequency domain spectrogram.}
\label{sec:co-channel}
\vspace{-0.9cm}
\end{center}
\end{figure}
Supervised approaches use model-based techniques to learn representations for both single speaker and overlapping speech segments. One of the most successful methods used for overlapping speech detection and separation is Non-negative Matrix Factorization (NMF) introduced in \cite{lee1999learning}. NMF is designed to learn the latent structure of the data by factorizing it into two non-negative matrices. One of these matrices is the basis matrix and the other one is the coefficients matrix. Next, the data is recovered by the weighted sum of the extracted basis. In our previous work \cite{yousefi2018assessing}, we used NMF to extract basis for every speaker in the dataset. Next, for each mixture, based on the extracted basis, we derived the coefficient matrix by minimizing the mean square error between the reconstructed mixture and the original mixture. The coefficient matrix gives us a lot of information regarding both the number of active speakers and the energy of each speaker in the mixture. Convolutive NMF (CNMF) \cite{yousefi2016supervised} is an extension for NMF that models the temporal continuity of speech signal by learning cross-column patterns as single basis which outperforms NMF in overlapping speech detection and separation. However, since NMF and CNMF are linear machine learning approaches, their abilities to cope with different types of overlapping speech in short segments of speech signals are limited. 

With the success of Deep Neural Networks (DNNs) in the past decade, several studies have applied different DNN architectures to address the classification of overlapping speech segments. One of the first studies in this area \cite{geiger2013detecting} uses Long Short-Term Memory (LSTM) network to address overlapping speech detection. Spectral kurtosis, spectral flux, harmonicity and MFCC are used to train the LSTM using AMI corpus \cite{carletta2005ami} which results in 76\% accuracy in detecting overlapping segments. Since AMI corpus is not balanced in terms of the ratio between the number of overlapping and non-overlapping speech samples, authors of \cite{andrei2017detecting} have used artificially generated overlapping speech with predefined Signal to Interference Ratio (SIR). They used FFT, MFCC and spectral envelope on time windows of 25, 100 and 500 ms to train a CNN network. The claimed accuracy on 25 ms frames was 74\% and on 500 ms segments was almost 80\%. Also, they reported Fscore, which is the harmonic mean of precision and recall, to be 72\% on the former frame length and 80\% on the longer time frames.

In this study we propose a new CNN architecture to address overlapping speech detection on frame level segments going as short as 25 ms. We also explore the effects of different features such as spectral magnitude, pyknogram, Mel Filter-Banks (MFB) and MFCC with its derivatives on the performance of the overlapping speech detection considering both the computation time and the classification measures. Our proposed system outperforms the systems introduced in \cite{carletta2005ami,andrei2017detecting} by 10\% in accuracy and 15\% in Fscore on 25 ms segments.

\begin{figure}
\begin{center}
\includegraphics[width=\linewidth]{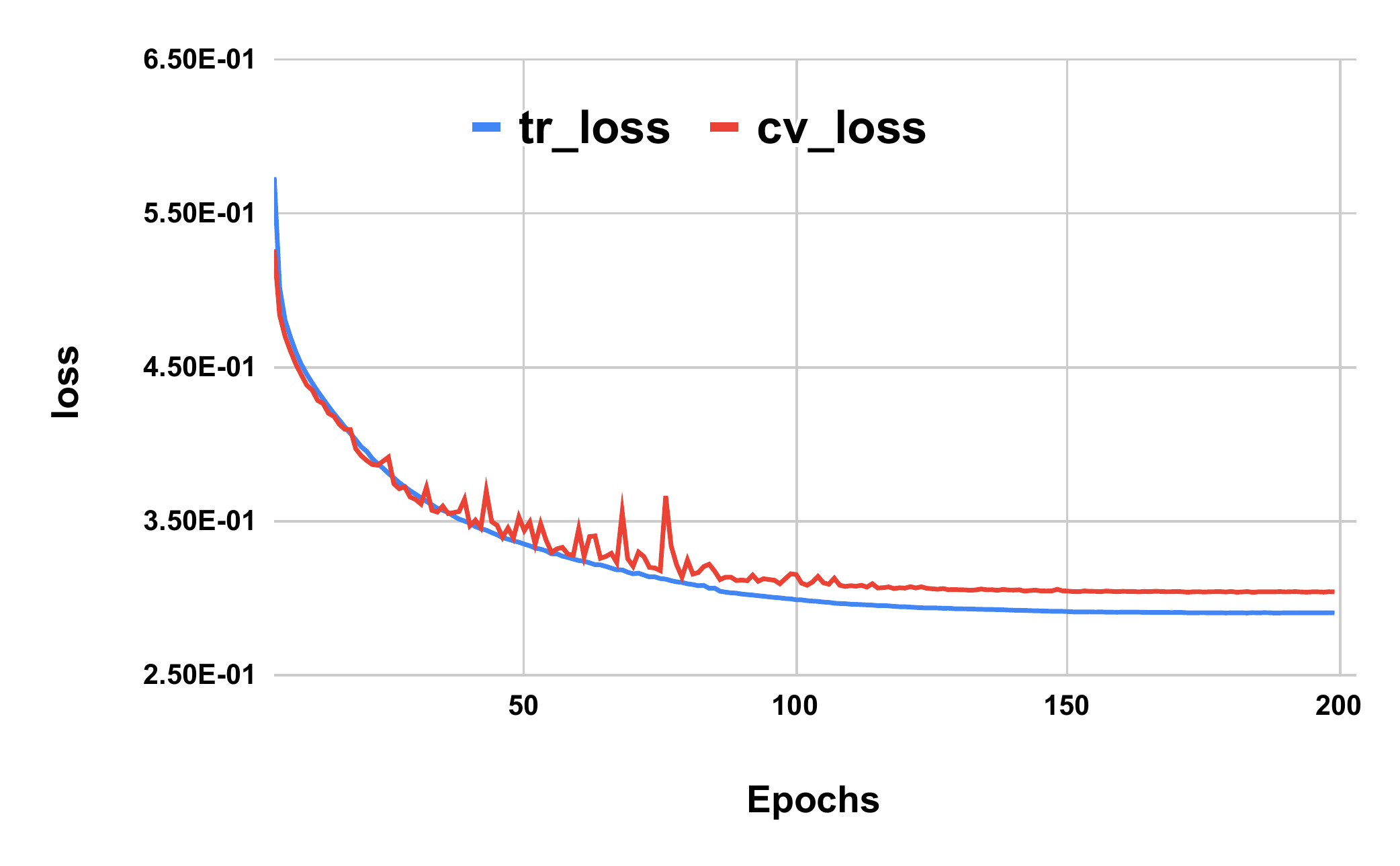}
 \caption{The training and cross validation loss for training the CNN model using Binary Cross Entropy (BCE) cost.}
\label{sec:loss}
\vspace{-0.5cm}
\end{center}
\end{figure}
\vspace{-0.5cm}
\section{Problem formulation}

As depicted in figure \ref{sec:co-channel}, the structure of speech signal is damaged in presence of the interfering talker. Therefore, different representations of speech i.e. spectrogram, harmonicity, and kurtosis can be used to detect an interfering talker. However, all these manually designed features have two drawbacks: first, they may not be the best representation for modeling competing talker, hence they lead to sub-optimal results; second, they can be fragile in noisy conditions. In order to extract optimal representations to model overlapping speech segments, Neural Networks(NNs) are trained based on well-designed dataset.

In \cite{boakye2008overlapped}, naturalistic data, such as AMI has been used to evaluate the overlapping speech detection system. However, as mentioned in \cite{von2019all}, AMI dataset contains only 5-10\% of overlapping speech, which is not sufficient for training DNNs. Also, the SIR of the overlapping segments in AMI corpus varies in short time frames, which makes it difficult for training a robust model. Hence, following the same approach used in \cite{andrei2017detecting}, we train our model based on artificially generated overlapping speech signals.

In this work we use a multi-speaker, sentence-based corpus called GRID, which has been used in monaural speech separation and recognition challenge \cite{cooke2006audio}. Also, this dataset has been used in several studies \cite{yousefi2018assessing,yousefi2019probabilistic} for overlapping speech detection and separation. This corpus contains 34 speakers, 16 female and 18 male speakers, each narrating 1000 sentence. For generating overlapping speech, random utterances, from random speakers are summed up with a random SIR, uniformly distributed between 0 and 5 dB. In order to make the generated mixture more similar  to  naturalist  data, the interfering  speech  is  added  to  the  target  speech from a random point.  Thus, each generated mixture file is either entirely overlapping speech or contains segments of both clean speech and overlapping speech. We have generated 20h of data for the training set, 3h for the development set and 2 hours for the test set. Also, the speakers used for generating the test set are not used in the training and development set.

\begin{table}
\begin{center}
\begin{tabular}{ |c|c|c|c|c| }
\hline
Male-Male &MagSpec&Pykno&MFB&MFCC\\
 \hline
Accuracy   & 79\% & 82\% & 78\%&81\% \\
\hline
Precision & 80\%   & 84\%& 81\%&82\%\\
\hline
Recall &    90\%& 91\%&91\%&90\% \\
\hline
Fscore & 85\%    & 87\%&86\%&86\% \\
\hline
Time & 898s  & 530s& 247s&220s \\
\hline
\end{tabular}
\caption{Evaluation of the proposed overlapping detection system on Male-Male overlapping speech signals. MagSpec is the spectral magnitude and Pykno is Pyknogram. Also the mean of processing time per epoch is reported in this table. }
\label{table:1}
\vspace{-0.6cm}
\end{center}
\end{table}

\vspace{-0.4cm}
\section{Model training and experiments}

Since the time domain waveforms are dense, using them directly for training the network is not computationally efficient. We extract a set of features itemized as below to train the network:
\vspace{-0.1cm}
\begin{itemize}
    \item 257-dim spectral magnitude.
    \vspace{-0.3cm}
    \item 40-dim Mel Filter-Bank (MFB).
    \vspace{-0.3cm}
    \item 39-dim Mel-Frequency Cepstral Coefficients (MFCCs) with their first and second derivatives.
    \vspace{-0.3cm}
    \item 120-dim pyknogram.
  
\end{itemize}
\vspace{-0.1cm}

Pyknogram is an enhanced version of speech spectrogram. It was shown in \cite{shokouhi2017teager} that pyknogram is more effective in detecting interfering speech than other features such as kurtosis, SAPVR and spectral flux. For extracting pyknogram, the speech signal is transformed into the spectro-temporal domain via gammatone filterbanks. Next, for each band-pass signal, the amplitude and frequency bin are computed using Teager-Kaiser Energy Operator (TEO) \cite{shokouhi2017teager}.  An advantage of TEO over conventional Fourier analysis is its capability in estimating energy in a nonlinear manner, which makes it a suitable tool for modeling speech signal. 

The sample frequency of the recordings are 8kHz. As the first feature, we used 512-dim magnitude spectra computed over a frame size of 25 ms with 10 ms of frame shift. In order to extract MFB features, we first apply pre-emphasis filter as $y(t) = x(t) - 0.97x(t-1) $ on the signal to amplify the high frequencies. Next, we calculate the STFT of the signals, then the energy of each frame is derived. A set of 40 triangular filterbanks are introduced to be applied on the energy of the frames. Finally, the logarithm of the output is considered as the MFB features. Computing the MFCC feature is the same as MFB with two extra steps: first, Discrete Cosine Transform (DCT) is applied on the log MFB to decorrelate the filterbanks coefficients; second, we apply sinusoidal liftering that gives less weight to the higher coefficients which provide less discrimination than the lower ones. The combination of 12 MFCC feature vector in addition to their first and second derivatives results in a 39-dim feature vector. For extracting pyknogram, the signal is passed through 120 gammatone filterbanks. Then the frequency bins found using TEO are compared to the bandwidth of its corresponding filter. If the frequency bin is within the bandwidth range, that bin is accepted, otherwise discarded. The pyknogram is a 120-dim feature vector per frame.

We use a CNN architecture to carry out the task of classifying segments of both overlapping and single talker speech. This architecture is similar to the successful "Deep speech 2" architecture introduced in \cite{amodei2016deep}. We tuned the hyper-parameters of the networks using the development set. The choice of 6 1-D convolutional layers with 128 output channels except for the final layer which has 32 output channel, is optimum. The kernel size of each layer is tuned to 2 and tanh activation function is applied to the outputs. The training phase is performed by completing 200 epochs with the batch size of 32. The network is updated by the the gradient of Binary Cross Entropy loss (BCEloss) using Stochastic Gradient Descent (SGD) with learning rate tuned to 0.001. Also, The learning rate is reduced by half if there is no improvement in cross validation loss for three successive epochs. The Training and cross validation loss of the network with the selected hyper-parameters are shown in figure \ref{sec:loss}, which depicts the ability of the network to generalize to the unseen speech segments in the development phase.
\begin{table}
\begin{center}
\begin{tabular}{ |c|c|c|c|c| }
\hline
Female-Female &MagSpec&Pykno&MFB&MFCC\\
 \hline
Accuracy   & 82\% & 84\% & 82\%&83\% \\
\hline
Precision & 83\%   & 86\%& 84\%&85\%\\
\hline
Recall &    91\%& 91\%&91\%&91\% \\
\hline
Fscore & 87\%    & 88\%&86\%&88\% \\
\hline
Time & 998s  & 536s& 250s&216s \\
\hline
\end{tabular}
\caption{Evaluation of the proposed overlapping detection system on Female-Female overlapping speech signals.}
\label{table:2}
\vspace{-0.6cm}
\end{center}
\end{table}
\begin{table}
\begin{center}
\begin{tabular}{ |c|c|c|c|c| }
\hline
Male-Female &MagSpec&Pykno&MFB&MFCC\\
 \hline
Accuracy   & 88\% & 89\% & 89\%&89\% \\
\hline
Precision & 91\%   & 92\%& 92\%&92\%\\
\hline
Recall &    91\%& 91\%&92\%&91\% \\
\hline
Fscore & 91\%    & 88\%&92\%&92\% \\
\hline
Time & 933s  & 510s& 230s&217s \\
\hline
\end{tabular}
\caption{Evaluation of the proposed overlapping detection system on Male-Female overlapping speech signals.}
\label{table:3}
\vspace{-0.6cm}
\end{center}
\end{table}
\begin{figure}
\begin{center}
\includegraphics[width=\linewidth]{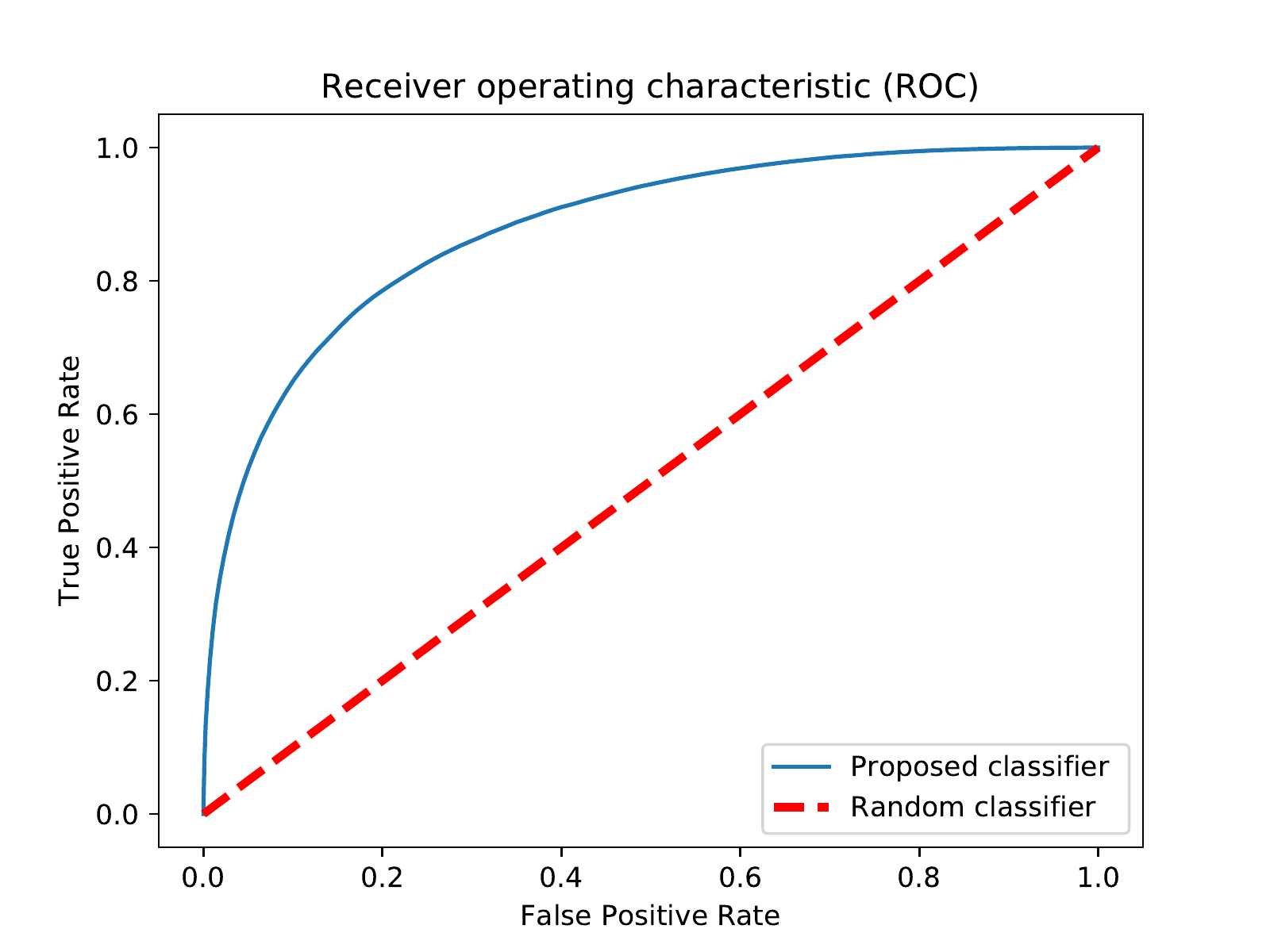}
\vspace{-0.7cm}
 \caption{The ROC curve of the proposed classifier for the pyknogram of the Male-Male mixtures.}
\label{sec:roc}
\vspace{-0.9cm}
\end{center}
\end{figure}

Since classification accuracy is not enough to evaluate the performance of the overlapping speech detection system, we also use other measures defined based on the confusion matrix such as precision, recall, and Fscore. Accuracy is the ratio between the number of the correct predictions divided by the total number of speech segments. For an unbalanced dataset, it is better to look closely to see how many examples have failed in each class. Therefore, precision and recall are frequently used to evaluate the performance of the system in unbalanced datasets. Precision expresses the ratio of the correctly detected overlapping segments to the total number of detected overlapping segments. However, recall is the ability of the model to find all the overlapping segments in the dataset which is measured as the ratio of correctly detected overlapping segments to the total number of actual overlapping segments. Fscore is another useful measure defined as the harmonic mean of recall and precision. Also, the processing time per epoch for each experiment is captured.

\begin{figure}
\begin{center}
\includegraphics[width=\linewidth]{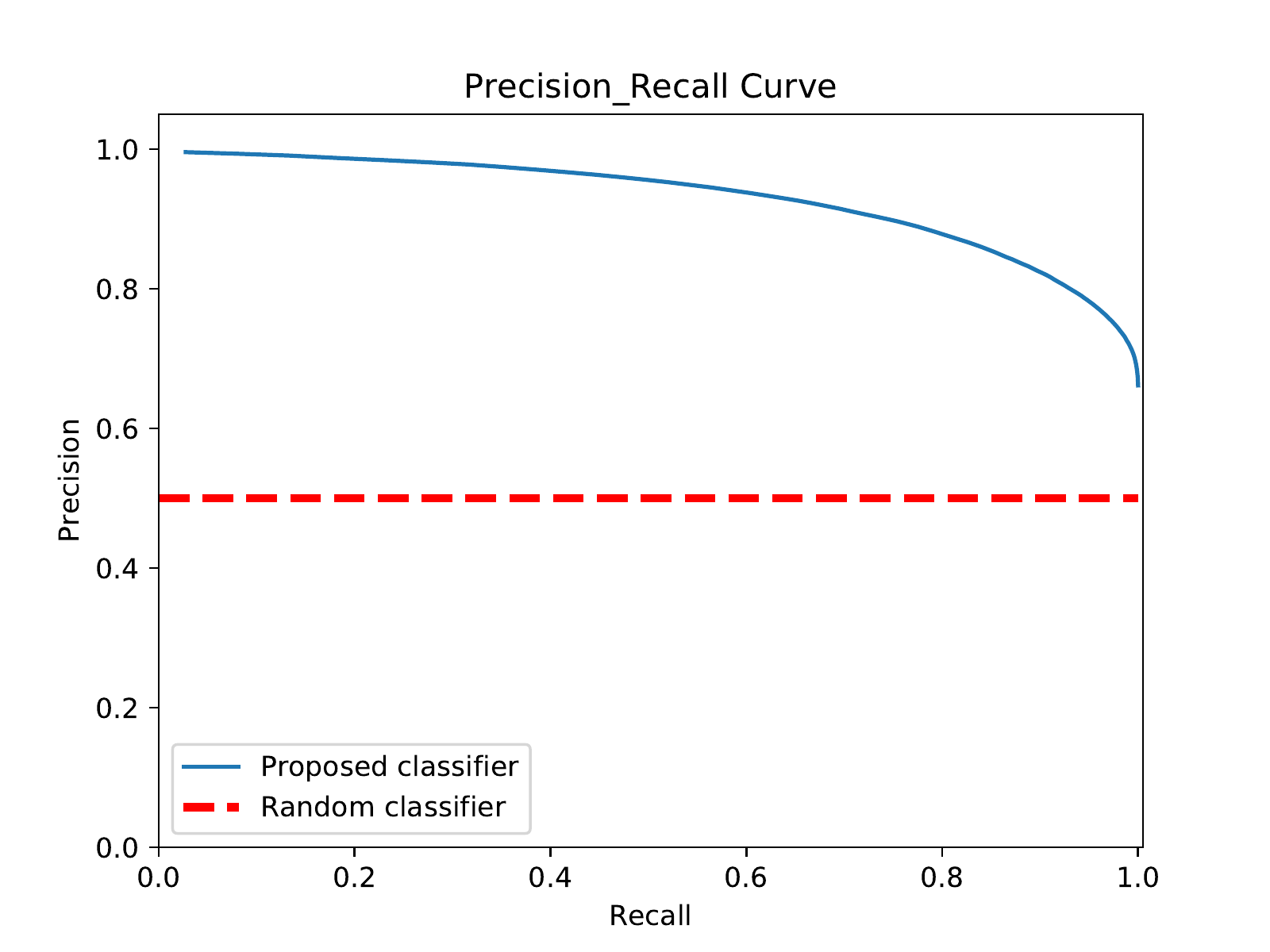}
\vspace{-0.7cm}
 \caption{The Precision-Recall curve of the proposed classifier for the pyknogram of the Male-Male mixtures.}
\label{sec:pr}
\vspace{-0.8cm}
\end{center}
\end{figure}

Tables \ref{table:1}, \ref{table:2}, \ref{table:3}  show the results of the experiments for three sets of data. In the first set, both target and interfering speakers are male, while in the second dataset, both speakers are female. 
The last set is generated by mixing male and female speakers. In the tables, MagSpec is the abbreviation for spectral magnitude and Pykno stands for pyknogram. As shown in table \ref{table:1}, the accuracy for spectral magnitude is 79\%, but since the test data is imbalanced in terms of class labels i.e. overlapping speech versus single speaker speech, other measures are better indicatives of the system performance.  Fscore of the male-male dataset is 85\% which generally manifests a good performance for the classification, however precision is 80\% and is 10\% lower than recall which is 90\%. Since magnitude spectra is a dense feature, the processing time is quite high in each epoch. The second largest feature is pyknogram, which outperforms spectrogram in both classification metrics and processing time. This manifests the capability of pyknogram in modeling the  structure of the speech signal. Therefore, it can easily detect the defects of the speech structure caused by the interfering talker. The other two features, MFB and MFCC also have higher accuracy and Fscore compared to spectral magnitude but less than pyknogram. However, since the dimensions of MFB and MFCC is lower than pyknogram (40 for MFB and 39 for MFCC), the processing time per epoch is much less than pyknogram. Another important aspect in choosing the right feature for overlapping speech detection is based on the application in hand. For those application where detecting all the overlapping segments are crucial while false alarms can be tolerated, a feature with higher recall is desired even if the precision and accuracy are low. For online applications, MFCC is the best option because the  required processing time and computational cost is the lowest compared to other features.  The performance for the female-female mixture set is shown in table \ref{table:2}, where the pattern of the results is almost the same as the male-male mixtures. This is expected and the slight difference in the results may be due to less challenging examples in the test dataset for the female-female mixtures. However, the results show that detecting overlap in the male-female mixture speech segments are far easier than the same-gender mixtures. This is due to the difference of the fundamental frequency of the speech signal for these two genders. Additionally, as a better demonstration of the classifier's performance, ROC and Precision-Recall curves are plotted in figure \ref{sec:roc} and \ref{sec:pr} in different operating points of the threshold, where the red line manifests the performance of a random classifier. These curves are based on the pyknogram of the male-male mixture dataset, which manifests a good performance for the proposed classifier.
\vspace{-0.5cm}
\section{Conclusion}
\vspace{-0.3cm}
We investigated the performance of different features for classifying overlapping speech segments in co-channel recordings. Spectral magnitude, pyknogram, MFCC and Mel Filter Bank (MFB) features are used to train a CNN architecture to tag overlapping segments. Pyknogram achieves the best performance, providing an accuracy of 84\% with Fscore of 88\% for the female-female overlapping speech. However, it is computationally less efficient than MFB and MFCC features. While the choice between these features will be dictated by the available compute, we demonstrate that using pyknogram can provide better classification accuracy in applications where online-processing is not required. 


\bibliographystyle{IEEEbib}
\bibliography{refs}
\end{document}